\documentclass[11pt,preprint]{aastex}

\begin{document}

\title{Resolving the Steep Spectrum Emission \\
in the Central Radio Source in ZwCl 0735.7+7421}

\altaffiltext{1}{Naval Research Laboratory, Code 7213, Washington, DC
  20375; aaron.cohen@nrl.navy.mil}
\altaffiltext{2}{Department of Astronomy, University of Virginia, 
P. O. Box 3818, Charlottesville, VA 22903; tclarke@virginia.edu}
\altaffiltext{3}{Istituto di Radioastronomia CNR, Via P. Gobetti, 101, 
I-40129 Bologna, Italy}

\author{A. S. Cohen\altaffilmark{1}, T. E. Clarke\altaffilmark{1,2},
  L. Feretti\altaffilmark{3} and N. E. Kassim\altaffilmark{1}}

\begin{abstract}
It has long been know that an extremely steep spectrum radio source lies a the center of ZwCl 0735.7+7421, a galaxy cluster with high X-ray luminosity and a cooling core.  In this Letter, we present VLA observations of this radio source at both 1425 and 325 MHz.  With a resolution below $21''$ (75 kpc) for both 1425 and 325 MHz, we show the morphology of the central source, and find that it is most likely a large (400 kpc) radio galaxy rather than diffuse cluster emission.  We estimate a steep spectral index of $\alpha_{325}^{1400} = -1.54$ for the core (although it may be contaminated by lobe emission), while the outer lobes are extremely steep objects, both with $\alpha_{325}^{1400} < -3.1$.  We also find evidence for restarted core activity in the form of a set of inner lobes oriented at a somewhat different angle from the outer lobes. A spectral analysis extending the frequency range down to 74 MHz appears to show a turnover at very low frequencies. Comparison of the minimum energy radio pressures with the average thermal pressure surrounding the radio cavities from \citet{mcnamara05} shows that the radio lobes appear to be roughly in pressure balance with the thermal gas.

\end{abstract}


\section{Introduction}

The hot X-ray emitting gas in clusters of galaxies cools through
thermal bremsstrahlung emission for which the cooling time is
$t_{cool} \propto T^{0.5}\, n_{\rm e}^{-1}$, where $T$ is the gas
temperature and $n_{\rm e}$ is the thermal electron density. In the
absence of a major cluster merger, the intracluster medium (ICM)
relaxes toward hydrostatic equilibrium. In the classical cooling flow
picture \citep{fabian94}, the relaxed cluster system develops an
inflow of gas toward the cluster center in order to maintain pressure
support against the cluster atmosphere. Early estimates of the mass
inflow rate from $Einstein$ and $ROSAT$ suggested cooling rates of 10
to 100's $M_\odot$ yr$^{-1}$ \citep{fabian94}. Recent X-ray
spectroscopic observations from $XMM-Newton$, however, show a greatly 
reduced amount of gas at intermediate X-ray temperatures (a few $10^6$K), 
and generally find that the gas is cooling
to only one-half to one-third of the ambient cluster temperature
\citep{peterson03}. To avoid the expected rapid cooling to very low
temperature, there must be some form of energy input into the
ICM to balance the cooling. Currently one of the best
candidates for supplying the energy is cluster-center radio
sources. \citet{birzan} show that in at least some clusters, the
energy input from the central radio source is sufficient to balance
cooling, at least for short time periods. 

We present an analysis of the radio properties of the exceptionally
steep-spectrum radio source (4C\,+74.13) in the core of the galaxy
cluster ZwCl 0735.7+7421. This cluster is a high X-ray luminosity
($L_x = 3.7\times10^{44}\ {\rm ergs\ s^{-1}}$
in the $0.2-2.5$ keV band; \citet{1995ApJ...449..554D}, adjusted for 
Wilkinson Microwave Anisotropy Probe [WMAP] parameters) 
system at the intermediate redshift
of $z = 0.216$. It was identified as a ``cooling flow'' candidate by
\citet{1992ApJ...385...49D} that was later confirmed by
\citet{1995ApJ...449..554D}. This cluster has been long known to
contain an extremely steep spectrum radio source at the center
\citep{1974MNRAS.168..307S}.  A 408 MHz image \citep{1974MNRAS.168..307S} 
hinted at extended structure but
did not have the resolution to characterize this, while a snapshot
image in the Very Large Array (VLA) A-configuration at 1.4 GHz showed 
only a very small ($\sim5''$) faint source \citep{1995ApJ...449..554D}.

In order to investigate the nature of the radio source in this unusual
system, we have conducted VLA observations at both 1425 and 325 MHz.
Throughout this Letter we use WMAP cosmological
parameters \citep{2003ApJS..148..175S} that give a comoving radial 
distance of 870 Mpc, and an angular scale of 3.47 kpc arcsec$^{-1}$.

\section{Observations}

We observed 4C\,+74.13 at 325 MHz with the VLA in B-configuration on
2003 December 4, and at 1425 MHz with the VLA C-configuration on 2004
April 15.  For
both frequencies, we used the multichannel continuum mode to reduce
the bandwidth smearing of bright sources throughout the field of view.
Also, both frequencies used 3C48 as a flux and bandpass calibrator.
Phase calibration was done using B0735+743 at 1425 MHz and B0739+703
at 325 MHz.  The entire primary beam area was imaged
in order to subtract confusing sources and the $uv$-data were weighted
with even balance between uniform and natural weighting (robust = 0).
Table \ref{tab1} summarizes the parameters of each data set.

The final image at 1425 MHz (Figure \ref{LC.fig}) has an rms noise level 
of 48 $\mu$Jy/beam and a synthesized beam size of $19.06'' \times 12.39''$ 
with PA = $-56.1^\circ$.
The final image at 325 MHz (Figure \ref{PB.fig}) has an noise level 
of 550 $\mu$Jy/beam and a synthesized beam size of $20.99'' \times 15.37''$ 
with PA = $13.07^\circ$.  

\section{Results}
\subsection{The 1425 MHz Morphology}

\begin{figure}
\plotone{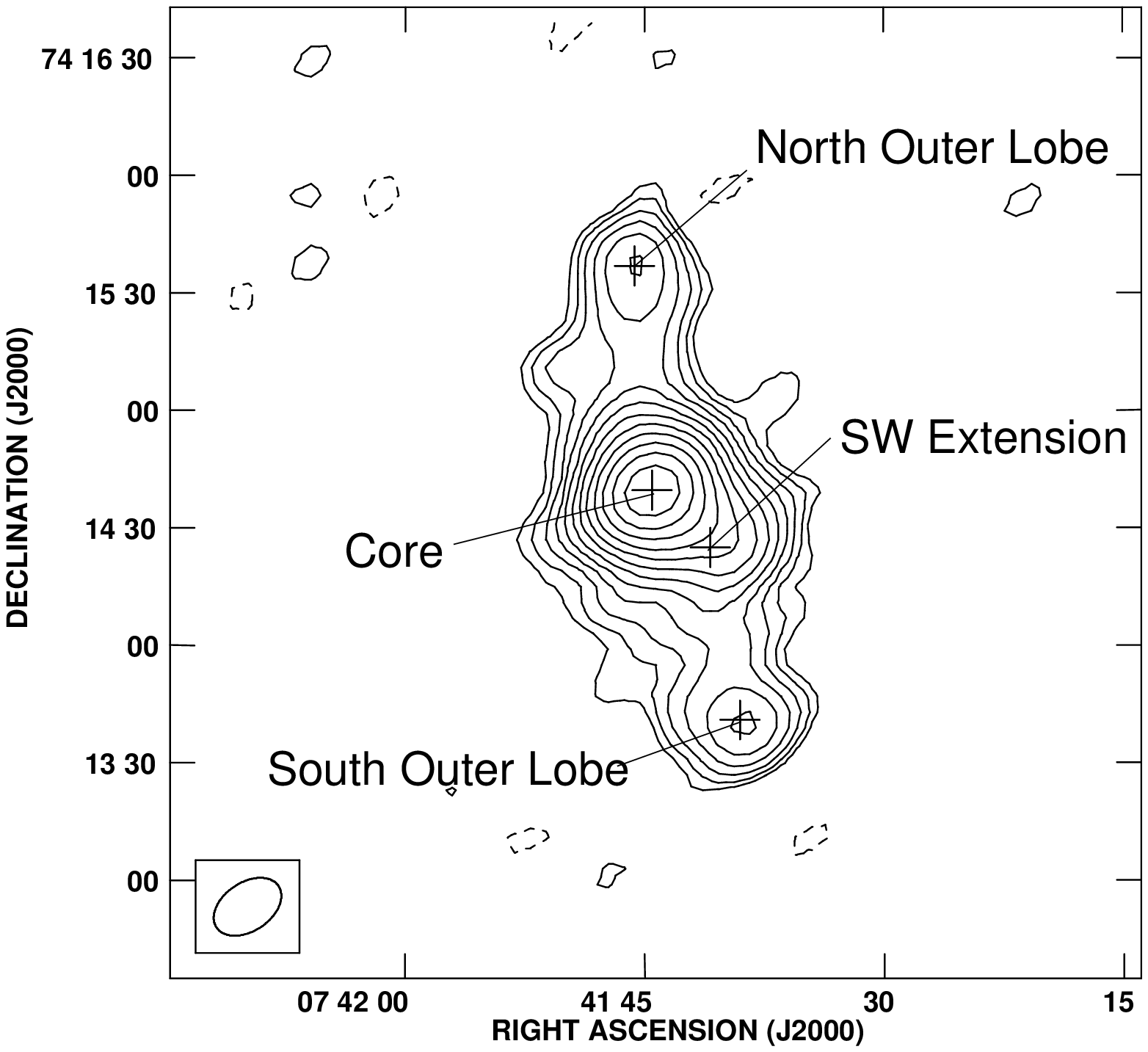}
\caption{Central radio source (4C\,+74.13) in the core of ZwCl
0735.7+7421 at 1425 MHz.  The contours begin at 120 $\mu$Jy/beam and
increase in factors of $\sqrt{2}$.  
}
\label{LC.fig}
\end{figure}

\begin{figure}
\plotone{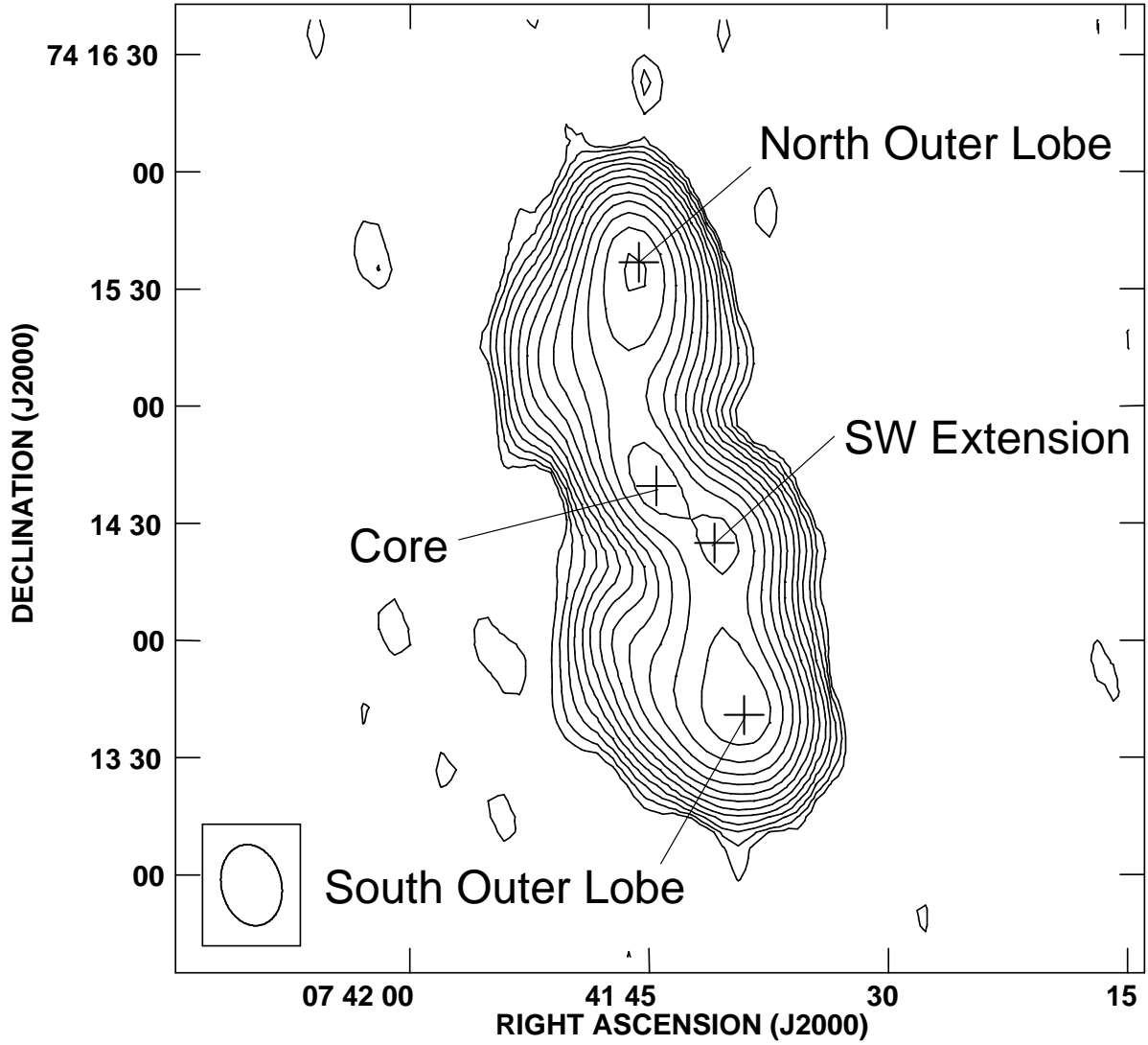}
\caption{Radio contours of the 325 MHz observations of 4C\,+74.13.
The contours begin at 1.5 mJy/beam and increase in factors of $\sqrt{2}$.  
This image is shown at the same scale as that in Figure \ref{LC.fig}.  
}
\label{PB.fig}
\end{figure}

\begin{figure}
\plotone{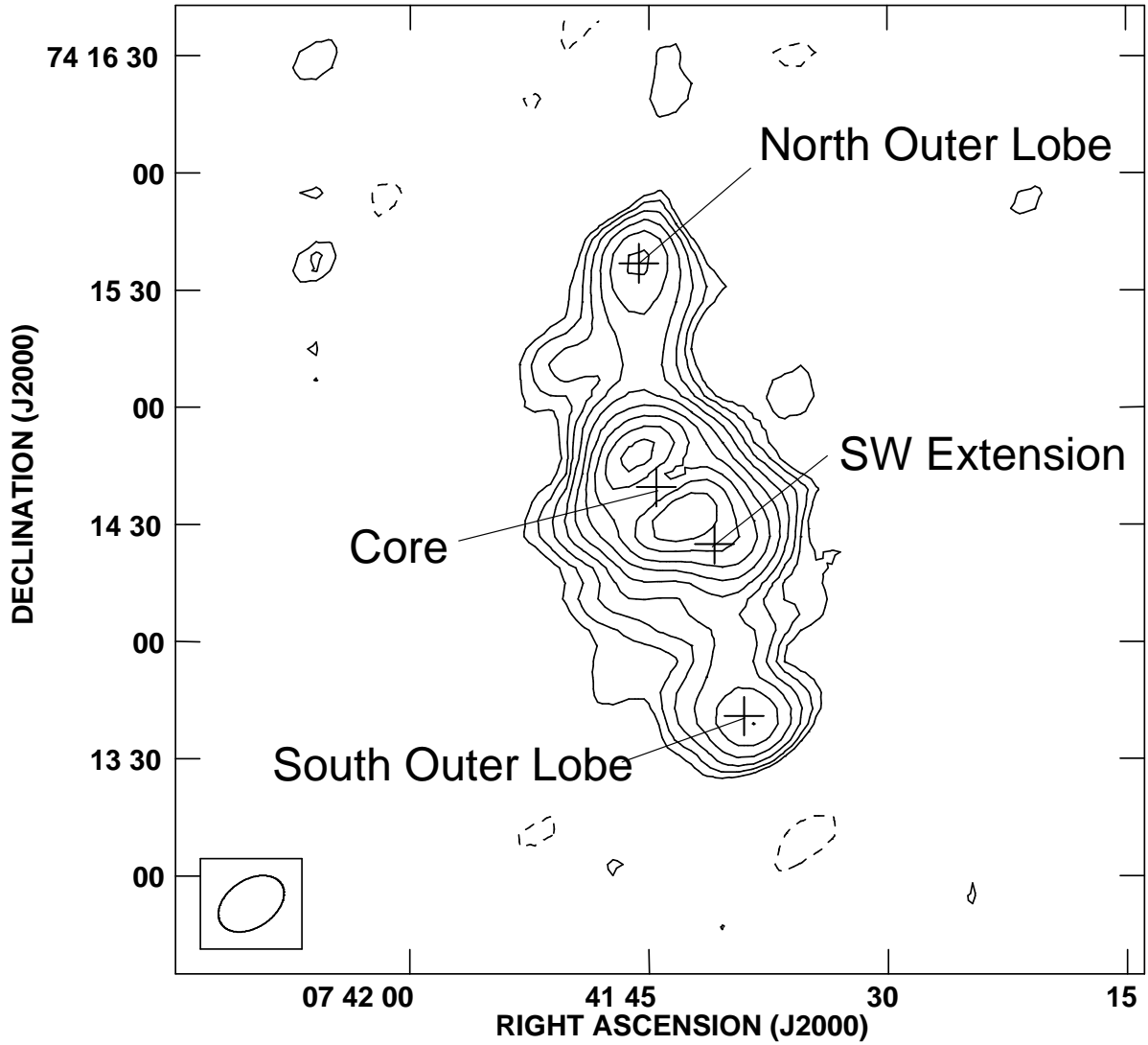}
\caption{Radio contours of the central region of ZwCl 0735.7+7421 at
1425 MHz with the core subtracted.  The contours begin at 120
$\mu$Jy/beam and increase in factors of $\sqrt{2}$.  This image is
shown at the same scale as that in Figure \ref{LC.fig}.
}
\label{LC.nocore.fig}
\end{figure}

\begin{figure}
\plotone{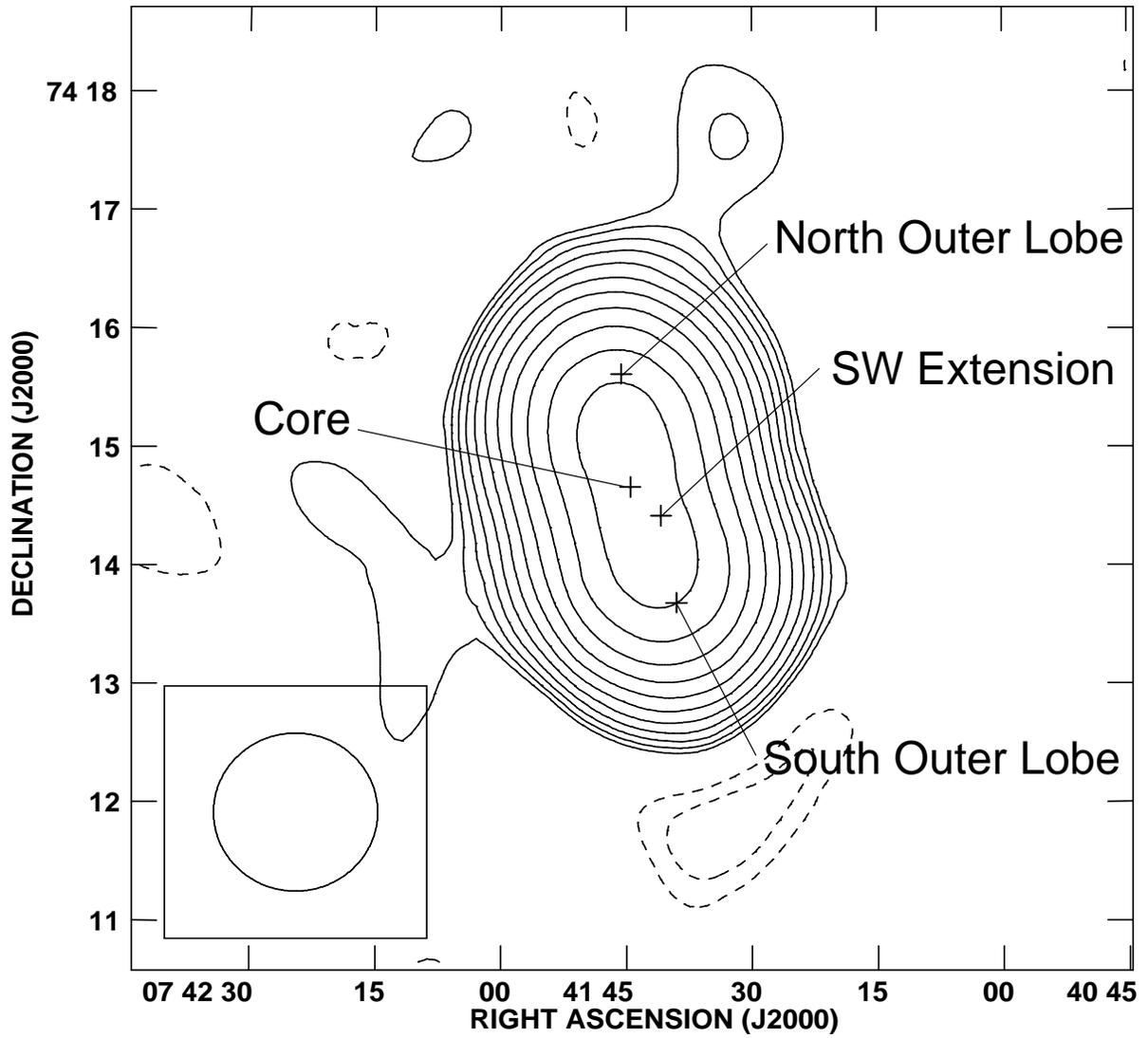}
\caption{ZwCl 0735.7+7421 at 74 MHz.  The contours begin at
150 mJy/beam and increase in factors of $\sqrt{2}$.  
}
\label{4B.fig}
\end{figure}

Our 1425 MHz image reveals a morphology consisting of three major peaks. 
The brightest peak is near the center and 
co-incident with the cD galaxy position listed by \citet{bauer00}, and
thus we take this to be the core of the radio galaxy.  The other two
peaks are consistent with radio lobes and are located roughly 58$''$
(200 kpc) to the north and south of the core. The lobes are not
co-linear with the core, possibly as a result of interactions with the ICM. We
determined the positions of these three peaks with single-Gaussian fits.  
These three positions are listed in Table \ref{tab2} and are labeled in Figures
\ref{LC.fig} - \ref{4B.fig}.

At the core, there is a significant extension of emission to the south
west, which we also label in Figures \ref{LC.fig} - \ref{4B.fig} as
the southwest extension.  This structure may be the result of the
radio jet interacting with the thermal ICM, or it may represent an
inner radio lobe from renewed activity in the core. Unfortunately, the
current radio data do not have sufficient resolution to separate these
possibilities. In Figure \ref{LC.nocore.fig} we show the 1425 MHz
image obtained after subtracting a point source located at the core
from the $u-v$ data.  This image reveals excess emission on both
sides of the core, suggestive of a set of inner radio lobes from
renewed activity.  Although higher resolution data are required to
confirm the possibility of a recent outburst, the data suggest that
either the jet axes have been deflected from their original directions
into the outer lobes, or the renewed activity has begun at a different
position angle from the original outburst that created the outer radio 
lobes.  We note that another cooling core cluster, A2199, contains the
famous renewed radio source 3C 338, which also has a steep spectrum.
\citep{1998ApJ...493..632G, 1998ApJ...493...73O}.

\subsection{The 325 MHz Morphology}

The 325 MHz image (Figure~\ref{PB.fig}) shows a similar overall extent
to the 1425 MHz image but significantly different morphology.  The core is
significantly fainter relative to the rest of the radio galaxy, indicating 
that the outer lobes are significantly steeper than the core. 
Also, the outer lobes are significantly more extended (particularly 
to the eastern edge) than at 1425 MHz. 
The eastern extension of the lobes at low frequencies may suggest relative
motions between 4C\,+74.13 and the ICM.

\subsection{Spectral Index Measurements}

We measure a total flux density for each frequency of $S_{325} = 820.0
\pm 41.1$ mJy and $S_{1425} = 21.94 \pm 0.54$ mJy.  These measurments 
are similar to the Westerbork Northern Sky Survey 
\citep{1997A&AS..124..259R} and NRAO VLA Sky Survey 
\citep{1998AJ....115.1693C} values of $S_{325} = 827.0$ mJy and 
$S_{1425} = 22.8$ mJy respectively.  The corresponding
spectral index ($S_\nu \propto \nu^\alpha$) of
$\alpha_{325}^{1425} = -2.45 \pm 0.04$ confirms
the extremely steep spectrum of this object.  In comparison, previous
observations of this source reported a spectral index of 
$\alpha^{1407}_{408} = -1.7$ \citep{1974MNRAS.168..307S}. 
A typical extragalactic radio source has a spectral index of $-0.7$
\citep{kk64} and a sample of 97 cluster radio sources has shown a typical
spectral index of $\alpha_{325}^{1400} \sim -1$ with the steepest at 
$\alpha_{325}^{1400} = -2.08$ \citep{2003MNRAS.339..913E}.

\begin{figure}
\plotone{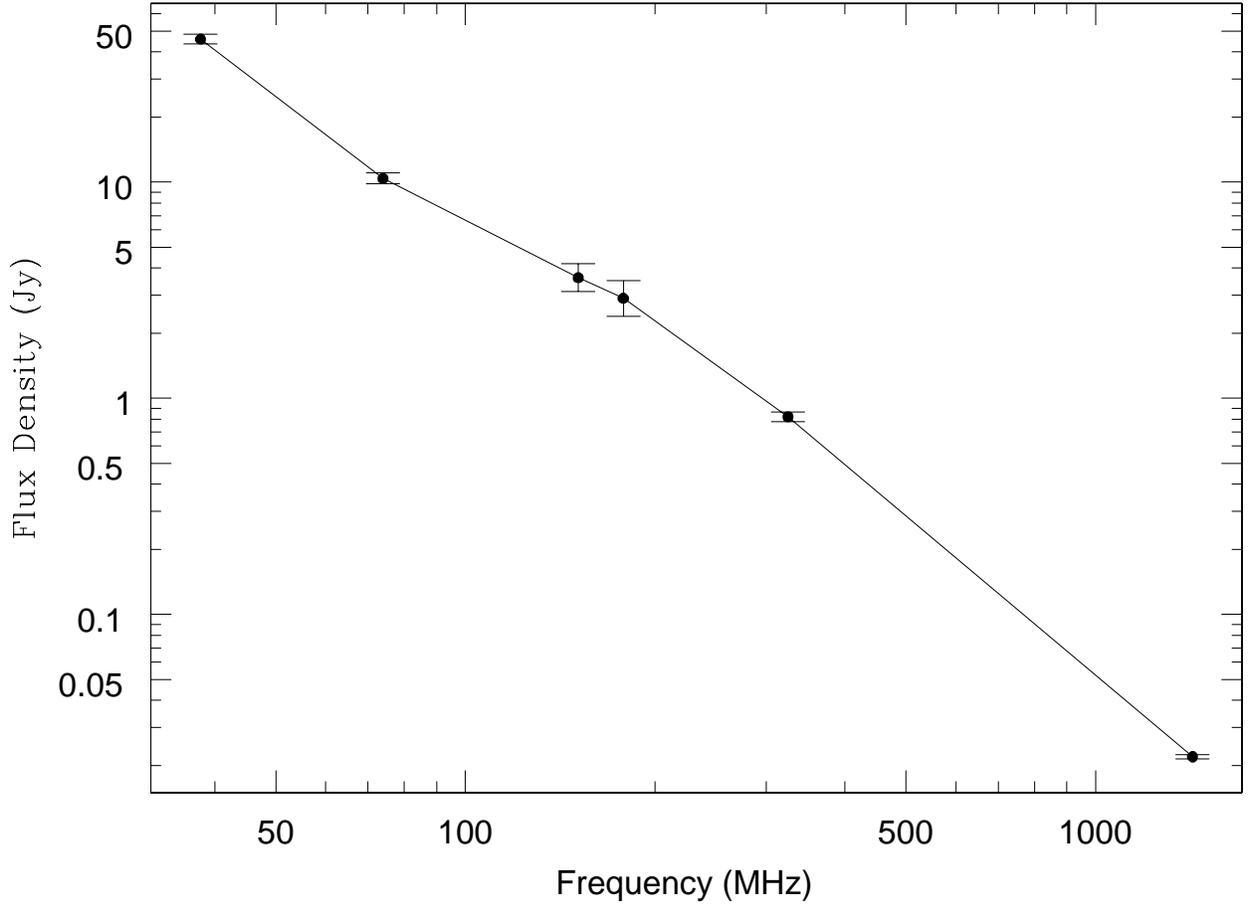}
\caption{Total flux density of 4C\,+74.13 as a function of frequency.
}
\label{spec.fig}
\end{figure}

We also measure the flux density at 74 MHz from an image
(Figure \ref{4B.fig}) taken from the VLA Low-Frequency Sky Survey
\citep[][]{2005CohenAAS}
to be $10.4 \pm 0.6$ Jy.  This implies a
spectral turnover toward low frequency with $\alpha_{74}^{325}
= -1.71 \pm 0.05$.  In Figure \ref{spec.fig} we plot these flux
measurements along with those from the 4C \citep[178
MHz;][]{1967MmRAS..71...49G}, 6C \citep[151
MHz;][]{1991MNRAS.251...46H} and 8C \citep[38
MHz;][]{1990MNRAS.244..233R} surveys.  While the 4C and 6C
measurements seem consistent with the overall trend of a spectral
turnover, the 8C measurement at 38 MHz is surprisingly high,
indicating that the spectrum turns up again ($\alpha_{38}^{74} =
-2.2 \pm 0.1$).  An examination of all images shows the contamination
of the 8C measurement to be unlikely, and so the low frequency spectral 
steepening appears real, although it needs to be confirmed
by more sensitive data.


Using the 4C flux measurement, we estimate the monochromatic radio
luminosity at 178 MHz to be $P_{178} = 5.2 \times 10^{26}$ W Hz$^{-1}$
sr$^{-1}$. This is roughly 4 times larger than the monochromatic power
of Hydra A \citep{taylor90}, and is typical of FR-II radio galaxies
\citep{FR}.

We have also measured the spectral indices of individual components of 
the radio galaxy.  First, we measured positions for these components by 
fitting a Gaussian to the region immediately surrounding the peak of each 
component.  For the core and the north and south outer lobes, we used 
the 1425 MHz image to measure the positions.  But, as the southwest 
extension is not a peak in the 1425 MHz image, we fitted the peak in the 
325 MHz image to determine its position.  We then convolved both images 
to a $21.0''$ circular beam and measured the surface brightness at the 
location of each component.  
The results are summarized in Table \ref{tab2}.  The outer
lobes have extremely steep spectral indices of $\alpha_{325}^{1425} =
-3.17 \pm 0.05$ and $-3.13 \pm 0.05$  for the north and
south outer lobes respectively.  
As discussed in Section 3.2, the outer lobes appear to show an extension
toward the east at 325 MHz that suggests spectral steepening in this 
region.  We have estimated the spectral indices on the eastern and western 
sides of each outer lobe and find a spectral gradient of $0.20\pm0.08$ for 
the north lobe and $0.15\pm0.08$ for the south lobe. This emission may 
represent a region of older radio plasma that is displaced to east by 
relative motions between the radio source and the thermal ICM.
The southwest extension has a somewhat
flatter spectral index of $-2.32 \pm 0.04$, which is
expected for a presumably younger particle population.  The spectral
index of the core is much flatter at $\alpha_{325}^{1425} = -1.54 \pm
0.04$.  As the flux density measurement for the core at 325 MHz is probably 
significantly contaminated by the surrounding lobe emission its spectral index 
is likely to be even flatter than this.

\subsection{Radio/X-Ray Comparison}

Recent $Chandra$ images of ZwCl 0735.7+7421 by \citet{mcnamara05}
reveal large ($\sim$ 200 kpc diameter) cavities in the thermal gas in
this system. They show that the 1.4 GHz radio emission of 4C\,+74.13
appears to fill the X-ray cavities. An X-ray temperature map shows that the
gas in the region of the cavities is significantly hotter than the
surrounding cluster, indicating that the radio outburst is driving
shocks into the ICM \citep{mcnamara05}. They estimate that a total energy
of $\sim 6 \times 10^{61}$ ergs was required to drive the shocks.

We have used our new radio observations to estimate the minimum-energy
synchrotron pressure in the outer lobes of 4C\,+74.13 for comparison
with the thermal pressure. We assume that the emission from the radio
lobes comes from a uniform prolate cylinder with a filling factor of
unity, and that there is equal energy in relativistic ions and
electrons. Using a model with the magnetic field perpendicular to the
line of sight, lower and upper cutoff frequencies of 10 MHz and
100 GHz, and spectral indices given in Table \ref{tab2}, we estimate 
minimum energy pressures of $3.9 \times
10^{-11}$ and $2.4 \times 10^{-11}$ dyn cm$^{-2}$ for the north and
south outer lobes, respectively. Based on the $Chandra$ data,
\citet{mcnamara05} estimate an average pressure surrounding the lobes
to be $P\simeq 6 \times 10^{-11}$ dyne cm$^{-2}$, roughly $1.5-2.5$
times higher than the minimum-energy pressures of the lobes. This
system appears to be much closer to pressure balance than most other
bubble systems that are generally an order of magnitude different
\citep{blanton04}. 

Using the spectral index of the full source ($\alpha_{325}^{1425} = -2.45$), 
we have calculated the total radio luminosity of 4C\,+74.13 over the
frequency range of 10 MHz - 10 GHz to be $L_{radio} = 38.2 \times 10^{42}$ 
ergs s$^{-1}$. Based on the spherical hydrodynamic shock
model, \citet{mcnamara05} estimate a required average radio jet power
of $P_s \simeq 1.7 \times 10^{46}$ ergs s$^{-1}$ for the radio
outburst that created the cavities. This suggests that the jet power
is a few hundred times the observed radio luminosity, similar to what
is found by \citet{birzan} for radio filled
cavities. \citet{mcnamara05} conclude that the radio outburst released
sufficient energy to balance the central cooling flow for several
gigayears. In comparison, \citet{birzan} found that roughly one-half of the
16 systems they studied had sufficient mechanical luminosities to
balance cooling in the cores.  
Although the \citet{birzan} sample specifically targetted 
clusters displaying X-ray surface brightness depressions, and thus the 
results are not generally representative of cluster systems,
we notice that the objects balancing the central cooling 
show a large spread in radio luminosity and in spectral index.

\section{Conclusions}

From these 1425 and 325 MHz data sets, we find that the central radio
source in the galaxy cluster ZwCl $0735.7+7421$ is an extended (400
kpc) radio galaxy with a clearly defined core and outer lobes.  The
overall spectrum is very steep between these two frequencies
($\alpha_{325}^{1425} = -2.45 \pm 0.04$ ) but turns over somewhat at
74 MHz with $\alpha_{74}^{325} = -1.71 \pm 0.05$, with an apparent
steepening to lower frequencies.  The spectral index of the core
($\alpha_{325}^{1425} = -1.54 \pm 0.04$) is very steep although much 
flatter than that of
the outer lobes, both with $\alpha_{325}^{1425} <  -3.1$.
We find evidence
of ``inner'' lobes, which may be the result of a recent radio
outburst.  The spectral index at the location of the southwest inner lobe
(the southwest extension of the core) is $-2.32 \pm 0.04$, which is
steeper than the core but flatter than the outer lobes.

Minimum energy estimates of the pressures of the radio lobes
are within a factor of 2.5 of the average thermal pressure in the
surrounding regions as determined by \citet{mcnamara05}. Using
estimates of the total radio luminosity, we find that the jet power
must be a few hundred times the observed radio luminosity.

Basic research in radio astronomy at the US Naval 
Research Laboratory is supported by the Office of Naval Research.
T. E. C. was supported in part by {\it Chandra} awards
GO2-3160X, GO3-4155X, GO3-4160X, GO4-5133X, GO4-5149X, and GO4-5150X
issued by the Chandra X-ray Observatory, which is operated by the
Smithsonian Astrophysical Observatory for and on behalf of NASA under
contract NAS8-39073.

{}

\begin{deluxetable}{lcc}
\tablecaption{Summary of Radio Observations\label{tab1}}
\tablehead{
\colhead{Observation:} & \colhead{90cm} & \colhead{20cm}
}
\label{obs.tab}
\startdata
Configuration & VLA B & VLA C \\
Central Frequency & 321.56 (IF 1) & 1385.1 (IF 1) \\
(MHz) & 328.50 (IF 2) & 1464.9 (IF 2) \\ 
Bandwidth (MHz) & 6.25 & 12.5 \\
Time on Source (min.) & 215 & 205 \\
Resolution ($''$) & $15.4\times21.0$ & $12.4\times19.1$ \\
RMS noise ($\mu$Jy/b) & 550 & 48 \\
\enddata
\end{deluxetable}

\begin{deluxetable}{llllll}
\tablecaption{Flux Densities and Spectral Index Measurements\label{tab2}}
\label{results.tab}
\tablehead{
\colhead{Component} & \colhead{$\alpha$ (J2000)} & \colhead{$\delta$ (J2000)} &
\colhead{$S_{1425}$} & \colhead{$S_{325}$} & \colhead{$\alpha_{325}^{1425}$} \\
 & \colhead{(h, m, s)} & \colhead{($^\circ$, $'$, $''$)} & 
\colhead{(mJy)} & \colhead{(mJy)} & \\
}
\startdata
Core & 07 41 44.54 & $+74$ 14 39.6 & $9.79 \pm 0.20$ & $94.9 \pm 4.8$ & $-1.54 \pm 0.04$ \\ 
S Inner Lobe & 07 41 40.88 & $+74$ 14 25.1 & $2.86 \pm 0.075$\tablenotemark{a} & $88.4 \pm 4.5$\tablenotemark{a} & $-2.32 \pm 0.04$ \\ 
N Outer Lobe & 07 41 45.63 & $+74$ 15 36.9 & $0.952 \pm 0.054$\tablenotemark{a} & $102.8 \pm 5.2$\tablenotemark{a} & $-3.17 \pm 0.05$ \\ 
S Outer Lobe & 07 41 39.02 & $+74$ 13 41.0 & $0.978 \pm 0.054$\tablenotemark{a} & $99.4 \pm 5.0$\tablenotemark{a} & $-3.13 \pm 0.05$ \\ 
Full Source & - & - & $21.94 \pm 0.54$ & $820.0 \pm 41.1$ & $-2.45 \pm 0.04$ \\ 
\enddata
\tablenotetext{a}{These values represent surface brightness in
mJy/beam, with circular beam FWHM = 21$''$.}

\end{deluxetable}

\end{document}